\documentclass[journal, 10pt]{IEEEtran}

\usepackage{subfigure}
\usepackage{pdfpages}
\usepackage{diagbox}
\usepackage{colortbl}
\definecolor{mygray}{gray}{0.85}
\usepackage{array}
\usepackage{caption}
\usepackage[linesnumbered,boxed,ruled,commentsnumbered]{algorithm2e}
        \usepackage{setspace}

 \usepackage{eqparbox}

\usepackage{amsmath,bm}
\usepackage{cite}
\usepackage{amsmath,amssymb,amsfonts}
\usepackage{algorithmic}
\usepackage{amsthm}
\usepackage{graphicx}
\usepackage{textcomp}
\usepackage{xcolor}

\usepackage{hyperref}
\hypersetup{%
  bookmarks=true
}

\usepackage{cite}
\usepackage{amsmath,amssymb,amsfonts}
\usepackage{algorithmic}
\usepackage{graphicx}
\usepackage{textcomp}
\usepackage{xcolor}
\usepackage{subfigure}
\usepackage{amsthm}
\usepackage{mathrsfs}

\usepackage{pdfpages}
\usepackage{colortbl}
\definecolor{mygray}{gray}{0.85}
\usepackage{array}
\usepackage{caption}
\usepackage{setspace}
\usepackage{diagbox}
\definecolor{mygray}{rgb}{0.96,0.99,0.96}
\definecolor{mypink}{rgb}{.99,.93,.85}
\definecolor{mycyan}{cmyk}{.32,0.04,0,0}
\usepackage[linesnumbered,boxed,ruled,commentsnumbered]{algorithm2e}
\usepackage{xcolor}  
\usepackage{colortbl,booktabs}
\usepackage{multirow}
\usepackage[ruled,linesnumbered]{algorithm2e}

{}
{}

\newcolumntype{I}{!{\vrule width 2.25pt}}
\newlength\savedwidth

\newlength\savewidth
\newcommand\shline{\noalign{\global\savewidth\arrayrulewidth
		\global\arrayrulewidth 1.350pt}%
	\hline
	\noalign{\global\arrayrulewidth\savewidth}}

\begin{document}
\title{\Huge
Spectrum Learning-Aided Reconfigurable Intelligent Surfaces for `Green' 6G Networks }

\author{
\IEEEauthorblockN{Bo~Yang,
 Xuelin Cao, Chongwen Huang, Yong Liang Guan, Chau Yuen, Marco Di Renzo,  Dusit Niyato,  M\'erouane Debbah, and Lajos Hanzo 
 }
\thanks{This paper has been accepted to be published by IEEE Network.}
 \thanks{B. Yang, X. Cao, and C. Yuen are with the Engineering Product Development Pillar, Singapore University of Technology and Design, Singapore 487372.} 
 \thanks{C. Huang is with College of Information Science and Electronic Engineering, Zhejiang University, Hangzhou 310027, China, and with International Joint Innovation Center, Zhejiang University, Haining 314400, China, and also with Zhejiang Provincial Key Laboratory of Info. Proc., Commun. \& Netw. (IPCAN), Hangzhou 310027, China.}
 
 \thanks{Y. L. Guan is with the School of Electrical and Electronic Engineering, Nanyang Technological University, Singapore.} 
 
\thanks{M. Di Renzo is with Universit\'e Paris-Saclay, CNRS, CentraleSup\'elec, Laboratoire des Signaux et Syst\`emes, 3 Rue Joliot-Curie, 91192 Gif-sur-Yvette, France.}
 
\thanks{D. Niyato is with the School of Computer Science and Engineering, Nanyang Technological University, Singapore.}

\thanks{M. Debbah is with the Lagrange Mathematical and Computing Research Center, 75007, Paris, and also with the Universit\'e Paris-Saclay, CentraleSup\'elec,  91190 Gif-sur-Yvette, France.}

\thanks{L. Hanzo is with the School of Electronics and Computer Science, University of Southampton, U.K.}
 }

\maketitle
 \begin{spacing}{1.0} 
\begin{abstract}
In the sixth-generation (6G) era, emerging large-scale computing based applications (for example processing enormous amounts of images in real-time in autonomous driving) tend to lead to excessive energy consumption for the end users, whose devices are usually energy-constrained. In this context, energy-efficiency becomes a critical challenge to be solved for harnessing these promising applications to realize	`green' 6G networks. As a remedy, reconfigurable intelligent surfaces (RIS) have been proposed for improving the energy efficiency by beneficially reconfiguring the wireless propagation environment.
In conventional RIS solutions, however, the received signal-to-interference-plus-noise ratio (SINR) sometimes may even become degraded. This is because the signals impinging upon an RIS are typically contaminated by interfering signals which are usually dynamic and unknown. To address this issue, `learning' the properties of the surrounding spectral environment is a promising solution, motivating the convergence of artificial intelligence and spectrum sensing, termed here as spectrum learning (SL). Inspired by this, we develop an SL-aided RIS framework for intelligently exploiting the inherent characteristics of the radio frequency (RF) spectrum for green 6G networks. Given the proposed framework, the RIS controller becomes capable of intelligently `{think-and-decide}' whether to reflect or not the incident signals. Therefore, the received SINR can be improved by dynamically configuring the binary ON-OFF status of the RIS elements. The energy-efficiency benefits attained are validated with the aid of a specific case study.  Finally, we conclude with a list of promising future research directions.
\end{abstract}

\section{Introduction}
The emerging sixth-generation (6G) network concept has to be ``green", while supporting ubiquitous artificial intelligence (AI)-assisted applications at a high energy-efficiency, including autonomous driving, unmanned aerial vehicle (UAV) aided logistics, rescue and surveillance, multi-access/mobile edge computing (MEC), and smart manufacturing~\cite{6G}. However, there are grave technical challenges ahead. Specifically, given the limited transmission powers, the received signal-to-interference-plus-noise ratio (SINR) will be considerably degraded due to severe signal fluctuations caused by both multipath fading and the presence of large obstacles. Although opting for millimeter wave (mm-Wave) and even Terahertz (THz) carrier frequencies has the potential of significantly improving the data rate, these frequencies inevitably suffer from high propagation losses, thereby they may result in an excessive energy consumption.

\begin{figure*}[t]
  \captionsetup{font={footnotesize }}
\centerline{ \includegraphics[width=6.25in, height=5.35in]{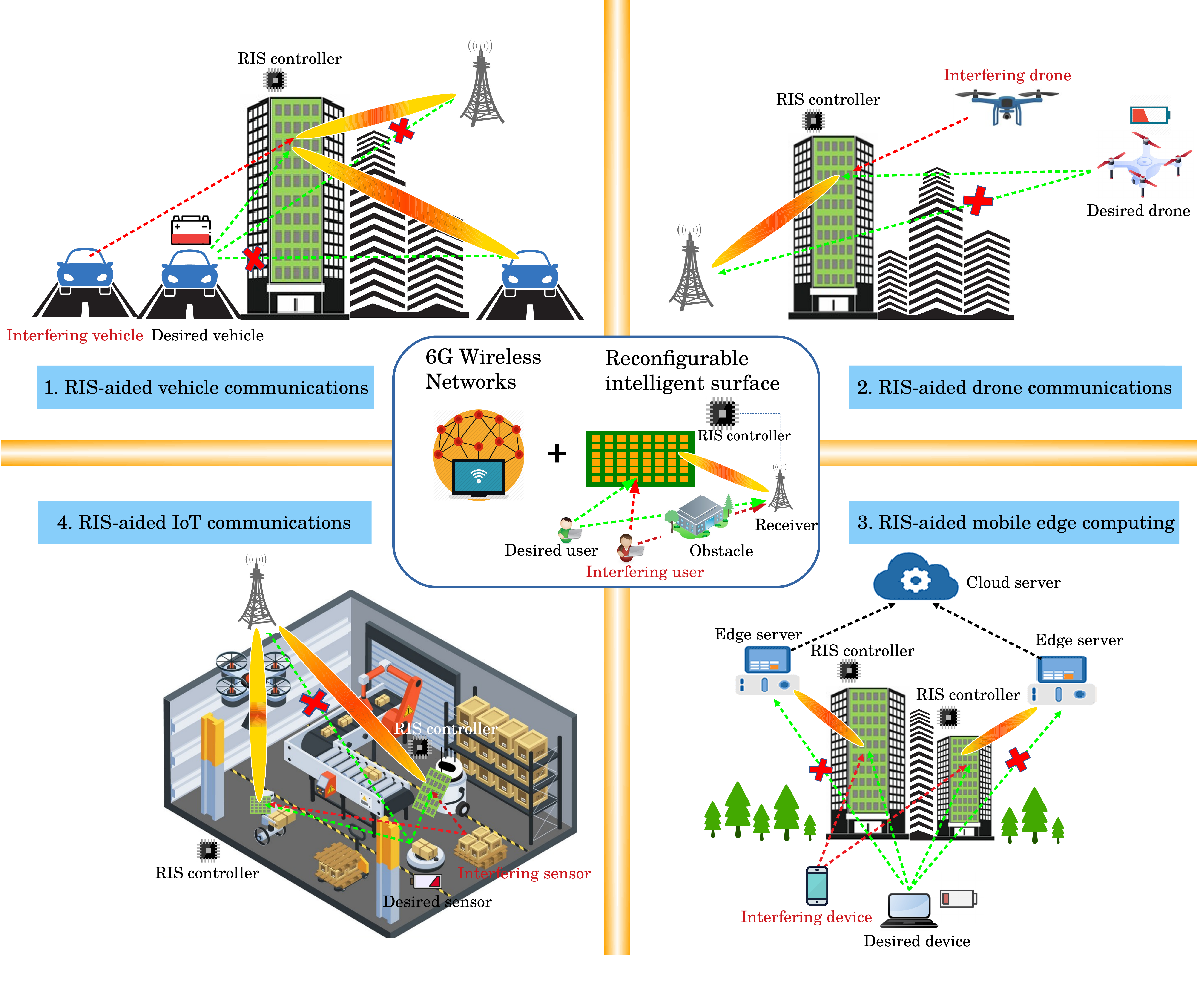}}
\caption{Emerging applications require RIS-aided green 6G wireless communications, including RIS-aided vehicle communications, RIS-aided drone communications, RIS-aided mobile edge computing, and RIS-aided IoT communications.}
\label{use_case}
\end{figure*}

In this context, reconfigurable intelligent surfaces (RIS) have emerged as a promising 6G solution for improving the quality of wireless links by appropriately reflecting the incident signals with the aid of a large number of nearly-passive reflecting elements at a low power consumption~\cite{6G_RIS}. Specifically, an array of passive scattering elements can be controlled in a software-defined manner to adjust the electromagnetic properties (such as the phase shifts and potentially the amplitude) of the incident radio frequency (RF) signals. As a result, the received signal strength can be enhanced via jointly optimizing the phase shifts of the reflecting elements of RISs~\cite{EURASIP_Editor2}. 
However, \textit{the interfering signals tend to dynamically fluctuate and a conventional RIS `blindly' reflects both the desired and interfering signals.}
 This may precipitate severe interference at the desired receiver and cause a deleterious effect. To address this challenge, it turns out that AI has a beneficial ameliorating potential in support of spectrum sensing, termed as spectrum learning (SL). This is achieved by intelligently identifying the characteristics of the wireless spectral environment. Hence, we envision that the convergence of SL and RIS will simplify network planning, paving the way for a green 6G ecosystem. 

In this article, we develop a dynamic SL-aided RIS framework for boosting the advantages of intelligent signal reflection through learning the inherent features of the incident RF signals at the RIS controller. The proposed framework `moves' the complexity of online detection to offline training, thereby controlling the binary ON-OFF status of RISs in real-time and improving the energy-efficiency of 6G networks.
The remainder of this article is outlined as follows. The following section provides an overview of green 6G services and its emerging applications empowered by RISs. Then, we present the design of the SL-aided RIS framework conceived for improving its energy-efficiency. Following that, we evaluate the performance of the proposed framework. Finally, we conclude this article with a list of compelling research opportunities.

\section{Reconfigurable Intelligent Surfaces Meet Green 6G Networks} \label{system_model}

\subsection{RIS-aided Green 6G Wireless Communication Systems}
Recent years have witnessed immense research efforts dedicated to RIS-aided wireless designs for improving the link quality and network coverage.  Sophisticated 3D index modulation, holographic MIMO schemes, mm-Wave and THz solutions, are expected to efficiently support high-quality services and seamless connectivity for a massive number of mobile terminals.
However, their excessive energy consumption and hardware costs pose design challenges due to the relatively high operating frequency, especially when many RF chains are required. Again, holographic MIMO surfaces attracted attention as a possible solution for realizing massive MIMO systems at a reduced cost and power consumption~\cite{Huang_WCM}. Notably, mobile devices are usually energy-constrained in practice, and the direct line-of-sight (LoS) link is often unavailable due to the occlusion of obstacles, when again, RISs come to rescue.

\begin{table*}[t]
\newcommand{\tabincell}[2]
		\centering
			\renewcommand{\arraystretch}{1.2}
			\captionsetup{font={small}} 
			\caption{Comparison of typical AI-empowered RIS research works.} 
			\label{RIS_AI}
			\small
			\centering 
			\begin{tabular}{|m{0.16\textwidth}<{\raggedright}|m{0.18\textwidth}<{\raggedleft }| m{0.2\textwidth}<{\raggedleft }|m{0.35\textwidth}<{\raggedleft}|  }   
				\shline
			    \rowcolor{mycyan}
			     \textbf{Objective}  & \textbf{Learning Approach} &\textbf{Model Architecture} & \textbf{Key Contributions}   \\
				\hline
			     { Channel estimation} & Deep learning~\cite{DL_CSI} & Multi-layer perceptron (MLP) with multiple layers & Jointly optimize the transmit beamforming matrix and the phase shifts together with only a few active elements on the RIS \\ \hline
			      {Joint beamforming and phase shift design} & Deep reinforcement learning~\cite{Huang_jsac} & Deep deterministic policy gradient (DDPG) with $4$-layered actor and critic networks & Jointly optimize the transmit beamforming matrix and the phase shifts \\ \hline
		 { Joint UAV trajectory and phase shift design} & Deep learning~\cite{Liu} &\textcolor{black}{Deep Q-network (DQN) with 3-layered convolutional neural network (CNN)} & Jointly optimize the movement of the UAV, phase shifts of the RIS, and power allocation
policy from the UAV to users.     \\ \hline
   Secure beamforming	& Reinforcement learning~\cite{DRL_security} & Model-free  & Jointly optimize the anti-jamming power allocation and reflecting beamforming without knowing the jamming model
\\		 \hline
			     Transmission strategy of each UAV-user pair   &  Multi-task learning~\cite{JSAC_CXL} &  Two-task (classification and regression) learning model with three hidden layers  & Jointly infer the optimal RIS elements allocation and RIS phase shifts of each UAV-user pair for energy-efficiency \\
			    \shline
			\end{tabular}  
	\end{table*}

As illustrated in Fig.~\ref{use_case}, an RIS-aided wireless communication system designed for energy efficient green 6G services typically supports a number of mobile devices. Each RIS element can be reconfigured by the RIS controller, which is connected to a central server (e.g., base station or access point) via a dedicated channel. 
To explore the potential of RIS-aided reflections, the following three steps are generally sequentially performed:

\begin{itemize}
\item \textit{Channel estimation:} In RIS-aided systems, there exist three kinds of wireless channels between the transmitter (Tx) and the receiver (Rx). By considering uplink transmission as an example, we have to estimate the wireless channels, including the Tx-RIS link(s), the RIS-Rx link(s), and the direct Tx-Rx link(s). To fully explore the potential gains brought about by RISs, the acquisition of accurate channel state information (CSI) becomes crucial, yet practically challenging.

\item \textit{RIS reconfiguration:} Following the acquisition of CSI, both the phase shift and the amplitude of RIS elements can be dynamically reconfigured, for example, based on positive-intrinsic-negative (PIN) diodes. This has been widely adopted in practical implementations as a benefit of its  low energy consumption and hardware cost.

\item \textit{Incident signal reflection:} After channel estimation and RIS reconfiguration, the RIS reflects the incident signal(s) with the optimized phase shift and amplitude. At the receiver, the direct and reflected signals are combined to improve the received
signal power.
\end{itemize}

\subsection{Emerging 6G Applications Empowered by RIS}
Conventional wireless communications are undergoing a significant transformation with the convergence of AI and 6G, evolving from the ``Internet-of-information" to the ``Internet-of-intelligence"~\cite{zy_iot}. This shall unleash the full potential of machine learning in innovative data-driven applications across a wide variety of industries, such as self-driving cars, UAV-aided surveillance, edge intelligence and smart factories. However, to support the aforementioned AI-based applications, huge amount of data is required for training. 
Against this background, we shall describe four typical RIS-empowered 6G applications, including terrestrial vehicular communications, drone communications, mobile edge computing, and IoT communications, as illustrated in Fig.~\ref{use_case}.  

\subsubsection{Autonomous Driving}
Video analytics relying on AI becomes an indispensable technique in support of autonomous driving, where a large amount of information generated by diverse onboard sensors (such as cameras, LiDAR sensors and radar sensors) has to be processed in near-real-time for collision avoidance in extremely complex surroundings.
However, promptly exploiting a deluge of delay-sensitive sensors' data  requires substantial onboard computation resources  and incurs high energy consumption~\cite{YB_magazine}. To maintain safety for self-driving, RISs usher in a paradigm shift from connected-vehicles to pervasive connected-intelligence by enhancing the link quality of cellular vehicular-to-everything (V2X) communications in complex urban environments at a low power.

\subsubsection{Unmanned Aerial Vehicles}
In recent years, UAVs, also commonly known as drones, have been extensively used for civilian, commercial, and military services in combination with AI~\cite{UAV_ZY}. However, the provision of sufficient energy for propulsion is a challenge, which may be mitigated by joint energy and data networking and laser charging~\cite{Hanzo}.
In this context, RISs show significant promise in terms of reducing the communications-related power of UAVs.

\subsubsection{Mobile/Multi-access Edge Computing}
At the time of writing, MEC is considered as a promising technique for offloading the processing of data to edge servers having much higher computing capabilities, thus enabling low-latency joint processing and communications. As a result, the AI processing is relocated from the central cloud to the network edge~\cite{iot_zy}.
 This implies that a large amount of data has to be transferred from the end devices to the edge server via band-limited uplink channels, which may become the bottleneck in MEC. In this context, RISs play a crucial role in mitigating the propagation-induced impairments with the aid of signal reflection.
 
\subsubsection{Smart Industry}
The smart industry, or ``Industry 4.0", integrating AI and the industrial IoT, paves the way for intelligent production by  interconnecting isolated industrial assets~\cite{iot_YB}. The wireless links are typically of non-line-of-sight (NLOS) nature, thereby leading to vulnerable links in dynamic environments. To circumvent this challenge, RISs improve the propagation environment by constructing signal reflections, thereby enhancing the robustness of wireless communications.

\begin{figure*}[t]
  \captionsetup{font={footnotesize }}
\centerline{ \includegraphics[width=5.65in, height=3.95in]{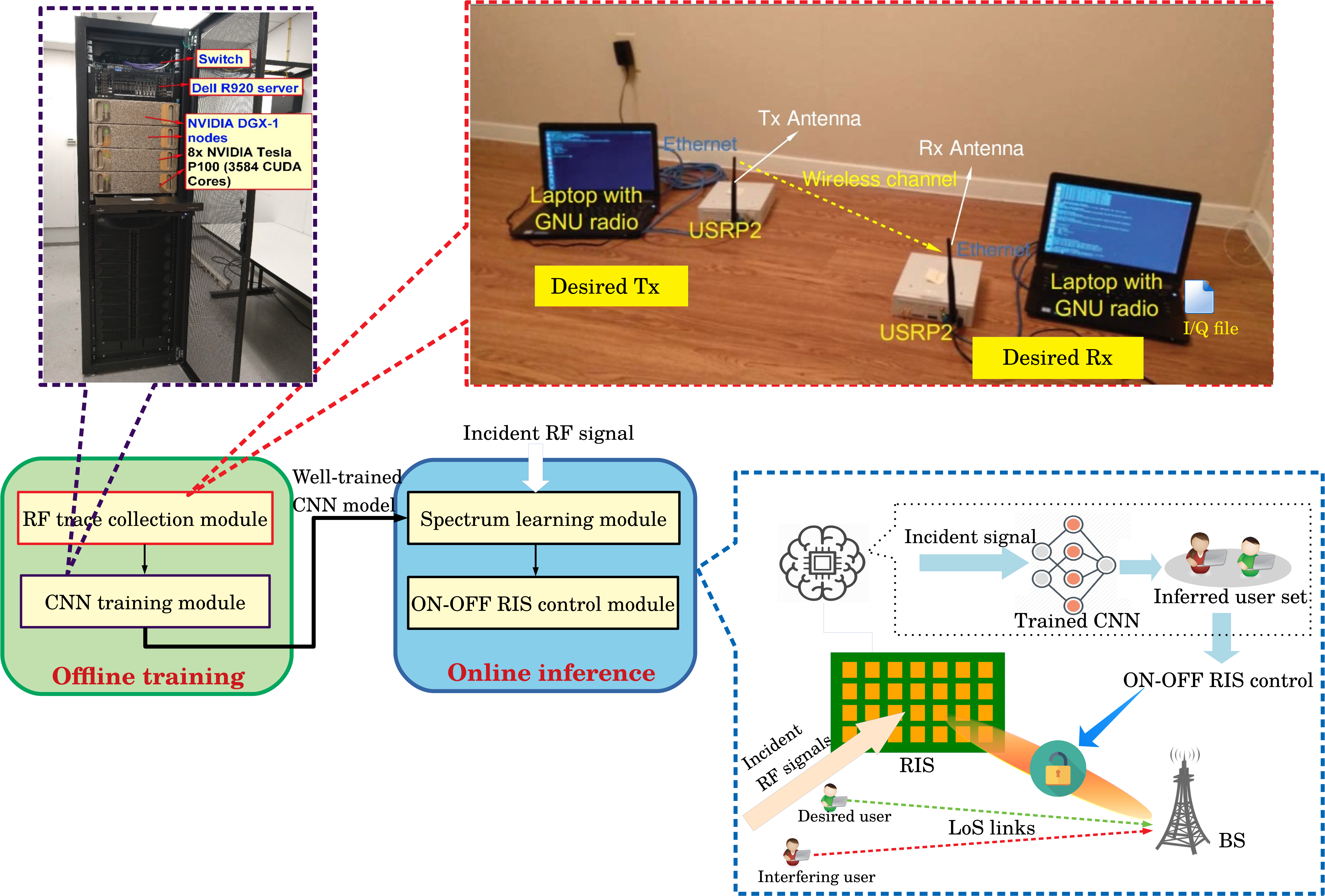}}
\caption{An illustration of the offline training stage and online inference stage in the proposed SL-aided RIS framework, where the RF trace collection module and the spectrum learning module are depicted.
}
\label{procedure}
\end{figure*}

\subsection{Challenge of Undesired-Reflections via RISs}
Despite their potential, RISs face numerous challenges, including the acquisition of accurate CSI, their reflection coefficients optimization, multi-user mobility management, and implementation issues, just to name a few. In particular, \textit{due to the unpredictable nature of interference, undesired reflections become a critical challenge, which has not been resolved yet.} Specifically, most of the existing treatises assume either that no interference exists, which rarely occurs in practice, or that the interference is already known. But naturally, it is not trivial to estimate the dynamically fluctuating interference in wireless networks. In reality, the SINR at the receiver might even be degraded compared to a system dispensing with RISs, unless specifically addressed. But the challenge is that RISs are nearly passive surfaces with no active sensing capabilities for channel and interference estimation. As a remedy, combining AI with spectrum sensing becomes attractive, which enables conventional RISs to learn the characteristics of the wireless environment. Hence, increasing research interests have been inspired by RIS-aided applications empowered by AI, which are contrasted in Table~\ref{RIS_AI}. 

In the following section, we shall present a dynamically configured ON-OFF\footnote{In this paper, the ON-OFF status is referred to having the entire set of elements of the RIS to be ON or OFF.} RIS control scheme based on spectrum learning, thereby enabling RISs to make intelligent decisions whether to reflect or not the incident signals, depending on the wireless spectrum environment inferred. Naturally, this intelligent action is only possible at the cost of incorporating intelligence into the RISs.

\section{Spectrum Learning-Aided RISs for Green 6G: Principles and Procedures}
\label{IIS}
\subsection{Design Principle} 
In a conventional system, the RISs passively and indiscriminately reflect all the incident RF signals. However, this kind of blind reflection sometimes contaminates the desired signal at the receiver. Fortunately, AI-aided spectrum sensing has the potential of intelligently orchestrating the incident signals by relying on an appropriately trained convolutional neural network (CNN) of the RIS controller.

More explicitly, the surrounding spectral environment is explored and learned by an intelligent classification algorithm, explicitly identifying the desired user, whose signal should be beneficially reflected, and the interfering users, whose signal should not be reflected. We arrange for having only a few active RIS elements for the baseband processing of the incident signal at the RIS controller at a low overhead\footnote{When the ratio of the active elements to the total number of elements is set to only $1\%$ and $7\%$ for a mmWave $28$ GHz scenario and for a low-frequency $3.5$ GHz scenario, respectively, the achievable rates can be maximized~\cite{DL_CSI}.}. \textcolor{black}{For simplicity, we assume that the base station (BS) of Fig.~\ref{procedure} has perfect CSI knowledge for all wireless channels, and that this CSI is fed back to the RIS controller via a dedicated control channel.} In particular, when an RF signal impinges upon an RIS, the RIS controller identifies the users that are included in the aggregate signal by using a well-trained CNN. Based on the inferred estimate, the RIS controller determines the ON-OFF status of the RIS for maximizing the SINR at the tagged BS. For example, the entire RIS is `turned OFF', if the received SINR at the tagged BS with respect to (wrt)  the RIS-aided link is lower than the corresponding SINR wrt the direct LOS link. Otherwise, the entire RIS is `turned ON'.

In the following, we will introduce the two stages of the proposed SL-aided RIS framework: the offline training stage and the online inference stage.

\subsection{Offline Training Stage}                                                                                                                                                                                                                                                                                                                                                                                                                                                                                                                                                                                                                                                                                                                                                                                                                                                                                                                                                                                                                                                                                                                                                                                                                                                                                                                                                                                                                                                                                                                                                                                                                                                                                                                                                                                                                                                                                                                                                                                                                                                                                                                                                                                                                                                                                                                                                                                                                                                                                                                                                                                                                                                                                                                                                                                                                                                                                                                                                                                                                                                                                                                                                                                                                                                                                                                                                                                                                                                                                                                                                                                                                                                                                                            
In the offline training stage, we have a pair of functional modules in support of spectrum learning.

\subsubsection{RF Trace Collection Module} 
  We collected RF traces by building a universal software radio peripheral (USRP2) based testbed, which is connected via the Gigabit Ethernet to a host computer, as illustrated in Fig.~~\ref{procedure}. \textcolor{black}{The RF trace collection relies on a transmit unit and a receive unit that communicate over a $1$ MHz channel at a carrier frequency of $2.4$ GHz. The traces are collected over the signal-to-noise ratio (SNR) range spanning from $0$ to $20$ dB. Since each USRP2 unit has one transmit antenna, multiple interfering signals are obtained by letting multiple USRP2 units transmit at the same time.} In particular, the transmit unit includes a transmitter host computer performing the baseband signal processing and a transmitter USRP2 for up-conversion, digital-to-analog (D/A) conversion, and wireless transmission. Upon receiving the RF signal via the wireless channel, the receiver USRP2 first processes the signals ahead from the radio interface and then performs A/D conversion and down-conversion.
 \textcolor{black}{Finally, the required baseband processing is carried out by the host computer, where the Inphase (I) and Quadrature (Q) sequences are stored for a wide range of SNRs to account for different interference scenarios. The acquired I/Q samples capturing the features of the incident RF signal are used for training the CNN model.}

  \subsubsection{CNN Training Module}                                                                                                                                                                                                                                                                                                                                                                                                                                                                                                                                                                                                                                                                                                                                                                                                                                                                                                                                                                                                                                                                                                                                                                                                                                                                                                                                                                                                                                                                                                                                                                                                                                                                                                                                                                                                                                                                                                                                                                                                                                                                                                                                                                                                                                                                                                                                                                                                                                                                                                                                                                                                                                                                                                                                                                                                                                                                                                                                                                                                                                                                                                                                                                                                                                                                                                                                                                                                                                                                                                                                                                                                                                                                                                                  
 In the proposed framework, spectrum learning is considered as a classification task, which is fulfilled via an appropriately trained CNN model. Specifically, \textcolor{black}{based on the acquired RF traces that amount to approximately one billion I and Q samples,} the CNN is trained offline based on a GPU cluster (e.g., the NVIDIA Tesla P100 GPU computing processor illustrated in Fig.~\ref{procedure}) by using the Adam algorithm that utilizes the cross-entropy as the loss function.
 \textcolor{black}{The CNN model trained consists of four layers: two convolutional layers that use the rectified linear unit activation function, two dense fully-connected layers and the output layer that uses the softmax activation function.}

\subsection{Online Inference Stage}
In the online inference stage, the RIS controller can dynamically adjust the ON-OFF status of the entire RIS to maximize the received SINR at the BS of Fig.~\ref{procedure} based on the appropriately trained CNN model.

\subsubsection{Spectrum Learning Module}
 
  \textcolor{black}{For the spectrum learning module, we can directly infer the spectrum access information related to the desired and interfering users by extracting the resultant I/Q samples from a copy of the impinging signals and feeding them into the appropriately trained CNN that is deployed at the RIS controller.} Specifically, as soon as the RF signal arrives at the RIS elements, it undergoes A/D conversion and frequency down-conversion. Then, baseband processing is executed and the extracted I/Q sequence is fed into the CNN model for spectrum access inference.
 
 \subsubsection{ON-OFF RIS Control Module} 
 The  ON-OFF RIS control module is trained for avoiding {undesired reflections} by dynamically setting the binary ON or OFF status of the RIS elements. In particular, we assume that both the CSI and phase shifts of the RIS can be perfectly estimated and optimized at the BS, which feeds this information back to the RIS controller via a dedicated  control channel\footnote{Channel estimation and phase shifts optimization of RIS have been widely investigated, hence they are not the main focus of this paper.}. By disabling and enabling the RIS (i.e., RIS is turned OFF and ON), the SINR that is achieved at the BS can be calculated respectively. If the calculated SINR with disabling the RIS is no worse than the corresponding SINR when the RIS is enabled, the entire RIS is turned OFF. Otherwise, the entire RIS is turned ON.

 \section{Case Study and Performance Evaluation}\label{results}
In this section, we characterize the experimental performance of the proposed SL-aided ON-OFF RIS control scheme.

\begin{figure}[t]
  \captionsetup{font={footnotesize }}
\centerline{ \includegraphics[width=2.5in, height=2.55in]{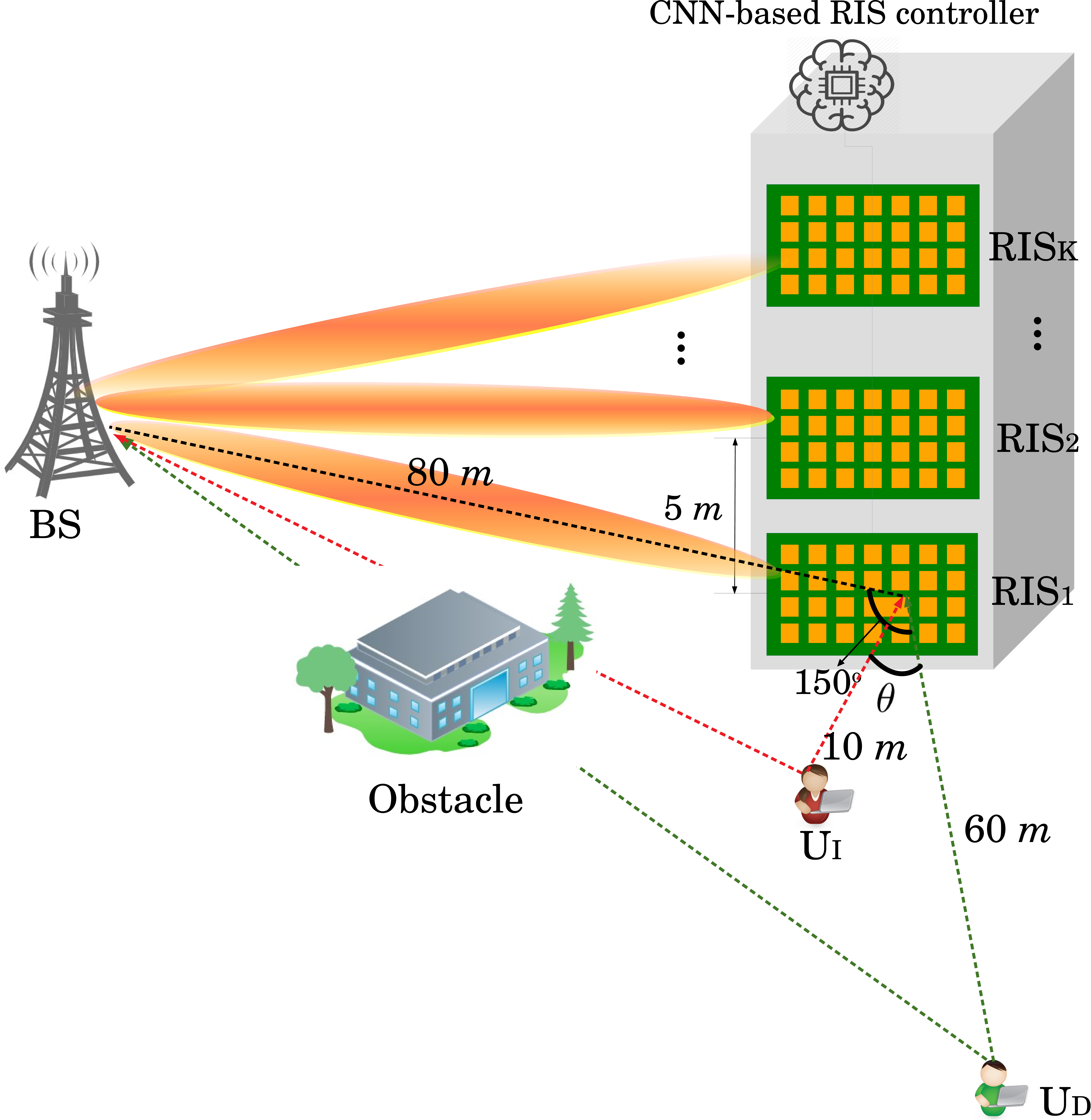}}
\caption{An experimental case of the proposed SL-aided RIS framework, where there exists an interfering user whose signal causes interference to the desired user's transmission at the BS. The angle of incidence between the desired user and the interfering user at the RIS is  $\theta$.
}
\label{CNN_solution}
\end{figure}

\subsection{Experimental Scenario}
For characterizing the proposed SL-aided RIS framework, we study the experimental scenario of Fig.~\ref{CNN_solution}, where the RIS reflects the signal arriving from both the {desired user} (denoted as $U_D$) as well as from the {interfering user} (denoted as $U_I$). As a result, the  signal received at the BS becomes a superposition of the desired signal from $U_D$ and the interfering signal from  $U_I$. 
The impact of the interference is, in particular, caused by undesired-reflections by the RIS and it is determined by the angle between the incident signal from the desired user and the incident signal from the interfering user (i.e., $\theta$ in Fig.~\ref{CNN_solution}). For example, the interference effect caused by $U_I$ at the BS may become more severe if $U_I$ is close to $U_D$, i.e., $\theta$ becomes small. In this case, the proposed SL-aided RIS framework enables the RIS controller to intelligently control the ON-OFF status of the RIS according to the wireless environment.


 In the scenario of two users ($U_D$ and $U_I$), we have four combinations of signals arriving at the RIS since each user has a binary state, i.e., active (also transmitting) or inactive (also idle): (1) ``Idle", indicating that both $U_D$ and $U_I$ are inactive; (2) ``\textit{$U_D$}", indicating that only $U_D$ is active, i.e., no interference occurs; (3) ``\textit{$U_I$}", indicating that only $U_I$ is active; and (4) ``$U_D\!+\!U_I$", indicating that both $U_D$ and $U_I$ are active. Upon receiving the superimposed incident signal, the RIS controller has to identify the signal's composition, which is a four-class classification problem readily solved by the trained CNN model\footnote{\textcolor{black}{Although a scenario with one interfering user is considered in this article for illustrative purposes, the proposed SL-aided RIS framework can be extended to more general scenarios where multiple interfering users coexist.}}.  \textcolor{black}{Therefore, the proposed SL-based framework `moves' the algorithmic and computational complexity to offline training while the complexity during the online inference is quadratic with time as a function of ${\cal O}(M^2 C K)$, where $C$ indicates the number of layers of the trained CNN model, $M$ is the number of neurons and $K$ indicates the total number of RISs.}

\subsection{Performance Evaluation}
In this part, we first analyze the inference accuracy of the CNN model with experimental data, and then we evaluate the performance of the proposed SL-aided ON-OFF RIS control scheme.

	\begin{figure}[t]
  \captionsetup{font={footnotesize }}
\centerline{ \includegraphics[width=3.1in, height=2.35in]{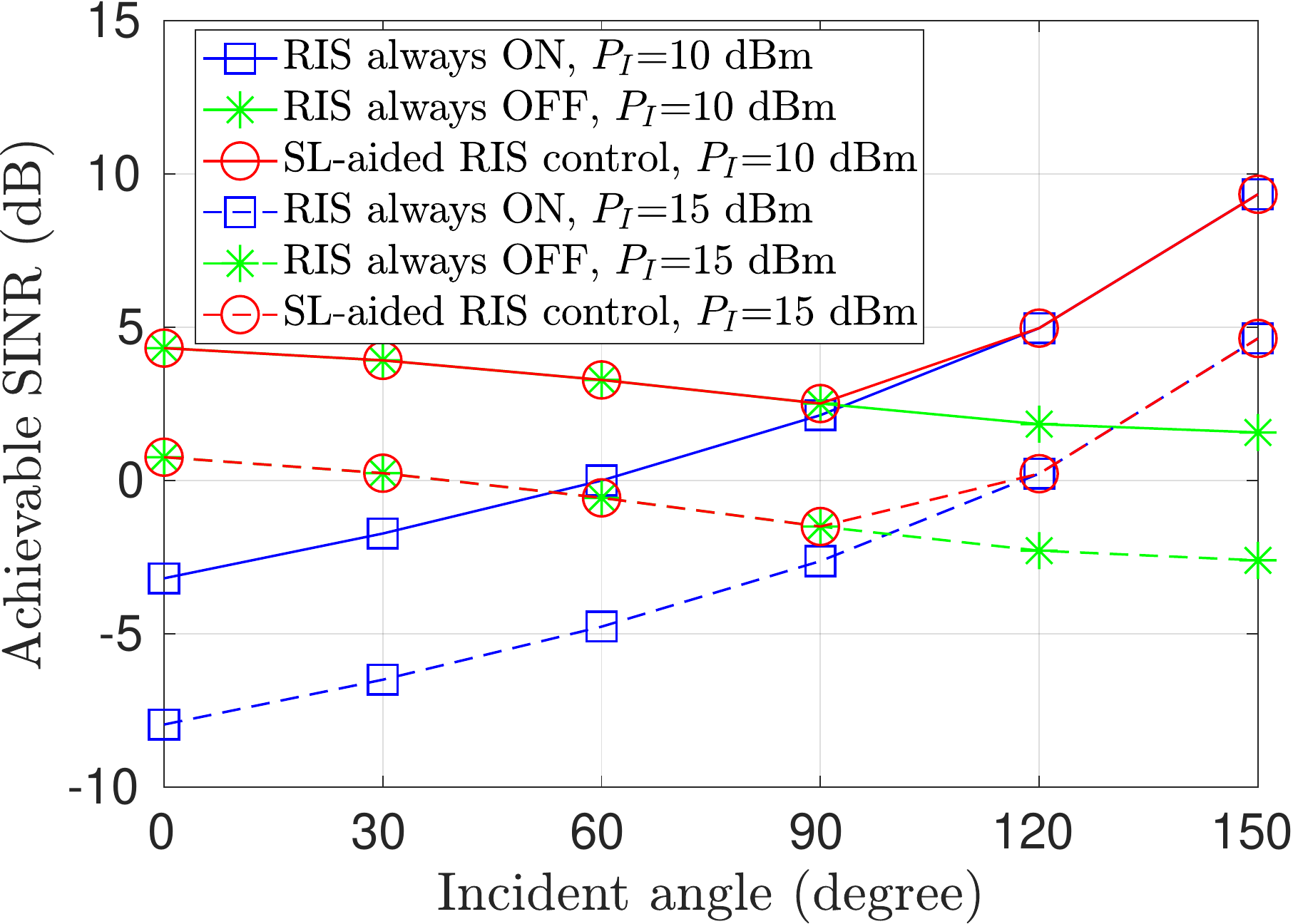}}

\caption{The achievable SINR versus the angle of incidence ($\theta$), where the number of RIS ($K$) is $1$. }
\label{result1}
\end{figure}

\subsubsection{Correct Inference Probability}
The correct spectrum inference probability of the trained CNN model is demonstrated by considering the experimental scenario of Fig.~\ref{procedure} supporting $U_D$ contaminated by $U_I$. During the CNN model training, the number of time series I/Q samples is set to $32$, $128$, and $512$. 
\textcolor{black}{The experimental results show that the correct inference probability obtained via the CNN model for the Class-2 (``$U_D$"), the Class-3 (``$U_I$") and the Class-4 (``$U_D\!+\!U_I$") is around $97\%$. Notably, the trained CNN model attains a perfect accuracy for the Class-1 (``Idle") prediction due to the {pattern distinguishable from the rest of the classes}.}  The relatively high inference probability of the trained CNN model results in typically feeding an accurate spectrum identification estimate to the  ON-OFF RIS control module. 
 

	\subsubsection{SL-enalbed ON-OFF RIS Control}
In the considered scenario, there is one desired user ($U_I$), one interfering user ($U_D$), and $K$ RISs. We assume that the RISs are equally spaced by $5 \ m$ in the vertical direction, each having $256$ elements. The  amplitude reflection coefficient is $1$ and the impact of interference is inversely proportional to $\theta$. As illustrated in Fig.~\ref{CNN_solution}, the RIS$_1$-$U_D$, RIS$_1$-$U_I$, and RIS$_1$-BS distances are $60 \ m$, $10\ m$, and $80 \ m$, respectively. Furthermore, the angle of incidence between the BS and $U_D$ at the RIS is $150\rm{^o}$, while between $U_D$ and $U_I$ is $\theta \in [0, 150\rm{^o}]$. To demonstrate the  energy-efficiency of the proposed SL-aided ON-OFF RIS control scheme, we evaluate a pair of benchmark schemes: 1) RIS always ON, and 2) RIS always OFF, where the transmission power of $U_D$ is $23$ dBm and the noise power is -$94$ dBm.
	
\begin{figure}[t]
  \captionsetup{font={footnotesize }}
\centerline{ \includegraphics[width=2.95in, height=2.35in]{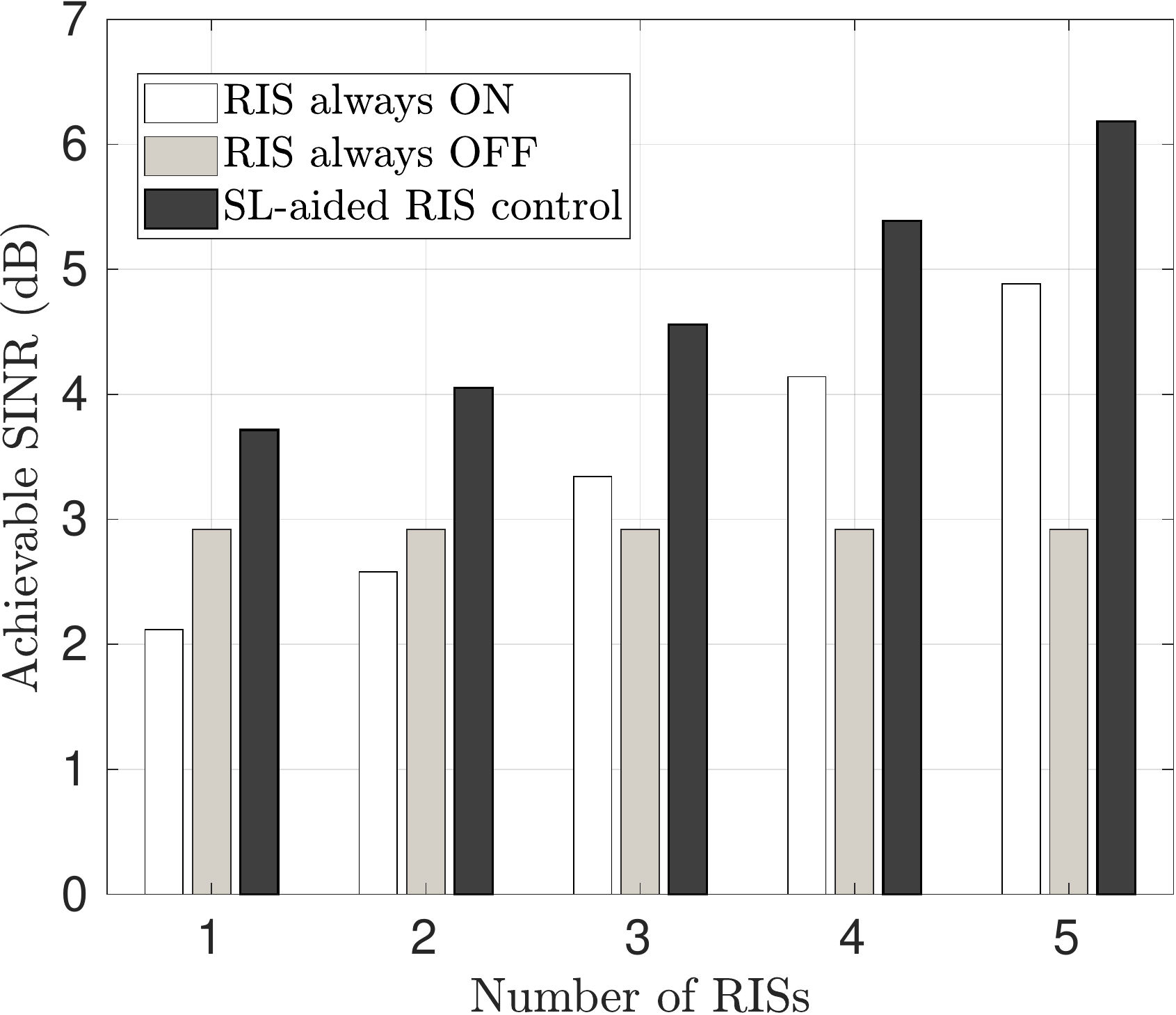}}
\caption{The achievable SINR versus the number of RISs, where $P_I=10$ dBm and $\theta$ is randomly selected from $[30\rm{^o}, 120\rm{^o}]$ considering $10^5$ realizations. }
\label{result2}
\end{figure}

The achievable SINR (in dB) versus the angle of incidence $\theta$ is depicted in Fig.~\ref{result1}, where a single RIS is considered, and the transmit power $P_I$ of the interfering user is $10$ dBm and $15$ dBm. We observe from Fig.~\ref{result1} that the benchmark schemes exhibit an opposite trend as $\theta$ increases. In particular, as $\theta$ increases, the interference contaminating the desired receiver through a reflection from the RIS is reduced. Therefore, it becomes more likely that the RIS is ON, thereby increasing the achievable SINR of the conventional `RIS always ON' scheme. By contrast, the achievable SINR of the `RIS always OFF' scheme reduces with $\theta$ due to the reduction of the distance between $U_I$ and BS. In this case, the interference reflected by the RIS becomes more pronounced.  \textcolor{black}{It can be observed that the SINR obtained by the proposed SL-aided RIS control scheme is consistently higher than that of the two benchmark schemes, which is the benefit of intelligently controlling the status of the RIS according to the dynamically fluctuating environment. The improvement of SINR via the proposed scheme is at the expense of increasing the RIS controller's complexity, i.e., the complexity introduced by the CNN online inference and the ON-OFF RIS state configuration. However, the associated complexity only grows linearly with the number of RISs, which is acceptable in practice.}

The impact of the number of RISs ($K$) on the achievable SINR is evaluated in Fig.~\ref{result2}, where we set $P_I=10$ dBm  and $\theta$ is randomly selected from $[30\rm{^o}, 120\rm{^o}]$. Intuitively, we observe that the SINR achieved by the `RIS always OFF' benchmark scheme remains unchanged, while that of the other two schemes increases with the number of RISs. Moreover, the SINR obtained by the proposed SL-aided RIS control scheme is significantly higher than that of the two benchmark schemes. In particular, when we use the same transmission power,  compared to the `RIS always OFF' and `RIS always ON' case studies, our proposed solution improves the received SINR from about $4.8$ dB and $2.9$ dB to $6.3$ dB for $K=5$. This demonstrates the potential of the proposed SL-aided RIS framework in terms of improving the energy efficiency of future 6G networks. 

\section{Conclusions and Promising Research Directions}
\label{conclusion}
In this article, we discussed the potential of spectrum-learning in addressing the critical challenges of RIS solutions in 6G. Explicitly, we  proposed an ON-OFF RIS control scheme for dynamically configuring the binary status of RISs. In this context, the beneficial role of SL in improving the  energy efficiency was validated by exemplifying the achievable SINR and was compared to a pair of benchmark schemes. 

Nonetheless, the proposed SL-aided RIS communication framework still remains largely unexplored, in particular, from its energy efficiency perspective. 
Achieving reliable SL necessitates the collection of a large amount of labeled RF sequences, and the appropriately trained model sometimes has to be fine-tuned or even retained, when the environment significantly changes, especially in vehicular and drone applications. This will lead to high energy consumption. 
To facilitate SL relying on a limited number of labeled RF sequences in a dramatically changing environment, advanced learning techniques, such as deep reinforcement learning, transfer learning and self-supervised learning may be explored in future works.



\end{spacing}


\begin{thebibliography}{00}
\bibitem{6G}
K. B. Letaief, W. Chen, Y. Shi, J. Zhang and Y. A. Zhang, ``The Roadmap to 6G: AI Empowered Wireless Networks," \textit{IEEE Communications Magazine}, vol. 57, no. 8, pp. 84-90, Aug. 2019.

\bibitem{6G_RIS}
C. Pan, H. Ren, K. Wang, M. Elkashlan, M. Chen, M. Di Renzo, Y. Hao, J. Wang, A.L. Swindlehurst, X. You, and L. Hanzo,
 ``Reconfigurable intelligent surface for 6G and beyond: Motivations, principles, applications, and research directions," \textit{arXiv preprint arXiv:2011.04300}, Nov. 2020.


\bibitem{EURASIP_Editor2}
M. Di Renzo, A. Zappone, M. Debbah, M. S. Alouini, C. Yuen, J. De Rosny, and S. Tretyakov, ``Smart radio environments empowered by reconfigurable intelligent surfaces: how it works, state of research, and the road ahead," \textit{IEEE Journal on Selected Areas in Communications}, vol. 38, no. 11, pp. 2450-2525, Nov. 2020.

\bibitem{Huang_WCM}
C. Huang, S. Hu, G. C. Alexandropoulos, A. Zappone, C. Yuen, R. Zhang, M. Di Renzo, and M. Debbah, 
``Holographic MIMO surfaces for 6G wireless networks: opportunities, challenges, and trends," \textit{IEEE Wireless Communications}, vol. 27, no. 5, pp. 118-125, Oct. 2020.

\bibitem{zy_iot}
Y. Wu, K. Zhang and Y. Zhang, ``Digital twin networks: a survey," \textit{IEEE Internet of Things Journal}, early access, doi: 10.1109/JIOT.2021.3079510.

\bibitem{YB_magazine}
B. Yang, X. Cao, K. Xiong, C. Yuen, Y. L. Guan, S. Leng, L. Qian, and Z. Han, ``Edge intelligence for autonomous driving in 6G wireless system: design challenges and solutions," \textit{IEEE Wireless Communications}, vol. 28, no. 2, pp. 40-47, Apr. 2021.

\bibitem{UAV_ZY}
K. Zhang, D. Si, W. Wang, J. Cao, and Y. Zhang, ``Transfer learning for distributed intelligence in aerial edge networks", \textit{IEEE Wireless Communications}, to be published.

\bibitem{Hanzo}
J. Hu, K. Yang, G. Wen and L. Hanzo, ``Integrated Data and Energy Communication Network: A Comprehensive Survey," \textit{IEEE Communications Surveys $\&$ Tutorials}, vol. 20, no. 4, pp. 3169-3219, Fourthquarter 2018.


\bibitem{iot_zy}
K. Zhang, J. Cao and Y. Zhang, ``Adaptive digital twin and multi-agent deep reinforcement learning for vehicular edge computing and networks," \textit{IEEE Transactions on Industrial Informatics}, early access, doi: 10.1109/TII.2021.3088407.

\bibitem{iot_YB}
B. Yang O. Fagbohungbe, X. Cao, C. Yuen, L. Qian, D. Niyato, and Y. Zhang, ``A joint energy and latency framework for transfer learning over 5G industrial edge networks," \textit{IEEE Transactions on Industrial Informatics}, early access, doi: 10.1109/TII.2021.3075444.

\bibitem{DL_CSI}
A. Taha, M. Alrabeiah and A. Alkhateeb, ``Enabling Large Intelligent Surfaces With Compressive Sensing and Deep Learning," \textit{IEEE Access}, vol. 9, pp. 44304-44321, Mar. 2021
 
\bibitem{Huang_jsac}
C. Huang, R. Mo and C. Yuen, ``Reconfigurable intelligent surface assisted multiuser MISO systems exploiting deep reinforcement learning," \textit{IEEE Journal on Selected Areas in Communications}, vol. 38, no. 8, pp. 1839-1850, Aug. 2020.

\bibitem{Liu}
X. Liu, Y. Liu and Y. Chen, ``Machine Learning Empowered Trajectory and Passive Beamforming Design in UAV-RIS Wireless Networks," \textit{IEEE Journal on Selected Areas in Communications}, vol. 39, no. 7, pp. 2042-2055, Jul. 2021.

\bibitem{DRL_security}
H. Yang, Z. Xiong, J. Zhao, D. Niyato, Q. Wu, H. V. Poor, and M. Tornatore,  ``Intelligent reflecting surface assisted anti-jamming communications: a fast reinforcement learning approach," \textit{IEEE Transactions on Wireless Communications}, vol. 20, no. 3, pp. 1963-1974, Mar. 2021.

\bibitem{JSAC_CXL}
X. Cao, B. Yang, C. Huang, C. Yuen, M. Di Renzo, D. Niyato, and Z. Han, ``Reconfigurable intelligent surface-assisted aerial-terrestrial communications via multi-task learning," \textit{IEEE Journal on Selected Areas in Communications}, early access,  doi: 10.1109/JSAC.2021.3088634.

\end{thebibliography}
\end{document}